%% file: main.tex
\documentclass[sigconf]{acmart}

\makeatletter
\def\@ACM@checkaffil{
    \if@ACM@instpresent\else
    \ClassWarningNoLine{\@classname}{No institution present for an affiliation}%
    \fi
    \if@ACM@citypresent\else
    \ClassWarningNoLine{\@classname}{No city present for an affiliation}%
    \fi
    \if@ACM@countrypresent\else
        \ClassWarningNoLine{\@classname}{No country present for an affiliation}%
    \fi
}
\makeatother

\usepackage[named]{algo}
\usepackage{amsfonts}
\usepackage{amsmath}
\usepackage{algorithm}
\usepackage{algorithmic}
\usepackage{ulem}
\usepackage{color}
\usepackage{verbatim}
\usepackage{multirow}
\usepackage{graphicx}
\usepackage{graphics}
\usepackage{balance}
\usepackage{mathtools}

\def\x{{\mathbf x}}

\def\R{\rm I\!R}

\def\myWidth{1.0}

\definecolor{light}{gray}{.9}

\newcommand{\Comment}[1]{}

\newcommand{\offset}{\textsc{Offset}\xspace}

\newcommand{\yohay}[1]{\noindent{\textcolor{blue}{\{{\bf Yohay:} \em #1\}}}}

\copyrightyear{2021}
\acmYear{2021}
\setcopyright{acmcopyright}\acmConference[CIKM '21]{Proceedings of the 30th ACM International Conference on Information and Knowledge Management}{November 1--5, 2021}{Virtual Event, QLD, Australia}
\acmBooktitle{Proceedings of the 30th ACM International Conference on Information and Knowledge Management (CIKM '21), November 1--5, 2021, Virtual Event, QLD, Australia}
\acmPrice{15.00}
\acmDOI{10.1145/3459637.3481958}
\acmISBN{978-1-4503-8446-9/21/11}

\begin{document}
\fancyhead{} 

\title{Unbiased Filtering Of Accidental Clicks in Verizon Media Native~Advertising}

\author{Yohay Kaplan}
\affiliation{Yahoo Research, Haifa, Israel}
\email{yohay@verizonmedia.com}
\author{Naama Krasne}
\affiliation{Yahoo Research, Haifa, Israel}
\email{naamah@verizonmedia.com}
\author{Alex Shtoff}
\affiliation{Yahoo Research, Haifa, Israel}
\email{alex.shtoff@verizonmedia.com}
\author{Oren Somekh}
\affiliation{Yahoo Research, Haifa, Israel}
\email{orens@verizonmedia.com}

\begin{abstract}
Verizon Media (VZM) native advertising is one of VZM largest and fastest growing businesses, reaching a run-rate of several hundred million USDs in the past year. Driving the VZM native models that are used to predict event probabilities, such as click and conversion probabilities, is \offset\ - a feature enhanced collaborative-filtering based event-prediction algorithm.
In this work we focus on the challenge of predicting click-through rates (CTR) when we are aware that some of the clicks have short dwell-time and are defined as accidental clicks. An accidental click implies little affinity between the user and the ad, so predicting that similar users will click on the ad is inaccurate. Therefore, it may be beneficial to remove clicks with dwell-time lower than a predefined threshold from the training set. However, we cannot ignore these positive events, as filtering these will cause the model to under predict. Previous approaches have tried to apply filtering and then adding corrective biases to the CTR predictions, but did not yield revenue lifts and therefore were not adopted. In this work, we present a new approach where the positive weight of the accidental clicks is distributed among all of the negative events (skips), based on their likelihood of causing accidental clicks, as predicted by an auxiliary model. These likelihoods are taken as the correct labels of the negative events, shifting our training from using only binary labels and adopting a binary cross-entropy loss function in our training process. After showing offline performance improvements, the modified model was tested online serving VZM native users, and provided $1.18$\% revenue lift over the production model which is agnostic to accidental clicks.
\end{abstract}



\maketitle

\section{Introduction}\label{sec:introduction}

\begin{figure}[!t]
\centering
\includegraphics[width=\myWidth\columnwidth]{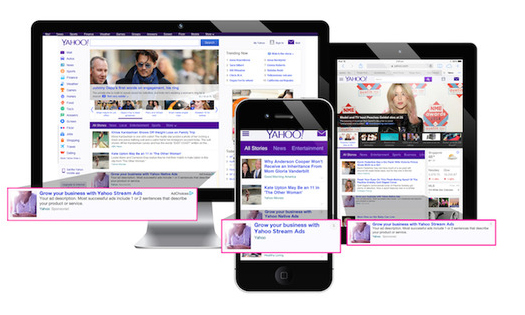}
\caption{Verizon media native ads on different devices.}
\label{fig:Gemini native}
\end{figure}

Verizon Media (VZM) native ad marketplace\footnote{See https://gemini.yahoo.com/advertiser/home} (previously known as \textit{Yahoo Gemini native}) serves users with native ads that are rendered to resemble the surrounding native content (see Figure \ref{fig:Gemini native}). In contrast to the search-ads marketplace, users' intent during page (or site) visits are generally unknown. Launched seven years ago and operating with a yearly run-rate of several hundred million USDs, VZM native is one of VZM main and fast growing businesses. With more than two billion impressions daily, and an inventory of a few hundred thousand active ads, this marketplace performs real-time \textit{generalized second price} (GSP) auctions that take into account ad targeting, and budget considerations.

In order to rank the native ads for the incoming users and their specific context according to the \textit{cost per click} (CPC) price type, a score (or expected revenue) is calculated by multiplying the advertisers' bid by the \textit{predicted click through rate} (pCTR) for each ad. Although VZM native handles other price types such as \textit{cost per video view} (CPV), in this work we focus on CPC price type. The pCTR is calculated using models that are continually updated by \offset\ - a feature enhanced \textit{collaborative-filtering} (CF) based event-prediction algorithm \cite{aharon2013off}. \offset is a one-pass algorithm that updates its latent factor model for every new mini-batch of logged data using a \textit{stochastic gradient descent} (SGD) based learning approach.

Currently, \offset treats all clicks as positive events, and all skips (impression with no clicks) as negative events. However, a closer look at click events which focuses on their \textit{dwell-time}\footnote{Roughly speaking, a dwell-time is the time a user spends at the landing page after clicking an ad.}, reveals that clicks can be classified into several types \cite{tolomei2019you}. In particular, \textit{accidental clicks} (AC) are clicks with relative short dwell-time (e.g., below $3$ seconds) indicating the user has probably clicked on the ad by accident. ACs are unwanted events since, on the one hand, they provide virtually no information about the affinity between users and ads, and consequently contaminate the model with wrong information. The model, in turn, is less capable of ranking ads based on the users' preferences, and therefore both revenue and user experience are negatively impacted. On the other hand, removing ACs from the training process would cause the click model to under-predict, or have a negative bias. Since advertisers are charged for all clicks, including accidental clicks, an under-predicting model will under-estimate the expected revenue, and due to the competitive nature of ad auctions it will unjustifiably loose opportunities to display ads to users. Consequently, we loose revenue, while the advertisers loose potential audience. It is concluded that ACs are undesirable events for all parties (users, advertisers, and publishers) and should be dealt with. However, reducing the number of ACs is \textbf{not} the objective of this work 

\Comment{We point out that charging advertisers for all clicks, including the accidental clicks, is a common business policy, and researchers developing scientific solutions for online advertising must operate within the constraints set by the business. We also note that reducing the number of accidental clicks, which may be a noble goal in its own right, is \textbf{not} the objective of this work.}

\Comment{[Original text] ACs are obviously unwanted events which provide poor experience for users and advertisers. Moreover, training a click prediction model on AC is  \yohay{'suspect'  instead of 'tricky?'}tricky since they \yohay{suggest replacing the rest of the sentence with '.. have more to do with site layouts than with any real tendency of the user towards the ad}relate more to the specific site layout rather than indicating a real tendency of the user towards the ad. On the other hand, removing ACs from the training process would cause the click model to under-predict which is a most unwanted phenomenon in online advertising systems. }

In this work we propose an approach for dealing with ACs in a way that improves our training process by filtering out ACs on one hand, while maintaining an unbiased click prediction model on the other.The unbiased AC filtering approach only requires changes in model training, without impacting other system components. In particular, we generalize the \offset algorithm to accommodate non-binary labels and use a binary cross-entropy loss function. While \textit{intentional clicks} (i.e., clicks with dwell-time above a certain threshold) are labeled with $1$s (positive events) as before, ACs and skips (negative events), are labeled with probabilities provided by an auxiliary AC prediction model instead of $0$s. As a result ACs are no longer ``contaminating'' the learning process and their positive weights are distributed among all the negative events. Therefore, the model is more accurate while remaining unbiased, in-spite of the AC filtering. After demonstrating improved prediction accuracy in offline evaluation, our approach was evaluated in online environment serving real VZM native traffic, showing significant revenue lifts over the AC agnostic production system.           
The main contributions of this work are:
\begin{itemize}
    \item Conducting a large scale analysis of click dwell-time that provides insights and supports various design decisions.
    \item Presenting a novel unbiased AC filtering approach to deal with the AC problem. The solution alters the click prediction model training and requires no changes in other system components. To our best knowledge, this is the first report of a successful AC aware web scale advertising system. 
    \item Performance evaluation in both offline and online settings, demonstrating a revenue lift of $1.18\%$ over the AC agnostic production system.
\end{itemize}

The rest of the paper is organized as follows. In Section~\ref{sec:background}, we provide relevant background, and discuss related work in Section~\ref{sec:related work}. After reporting dwell-time analysis findings in Section \ref{sec: Click Dwell-Time Analysis}, the problem at hand is defined and discussed in Section \ref{sec:problemDef}. Our approach is presented in Section \ref{sec:our approach}. Performance evaluation of the proposed approach is considered in Section \ref{sec:eval}. We conclude and discuss future work in Section~\ref{sec:conclusions}.

\section{Background}\label{sec:background}
\subsection{Verizon Media Native}\label{sec: Gemini Native}
\begin{figure}[!t]
\centering
\includegraphics[width=\myWidth\columnwidth]{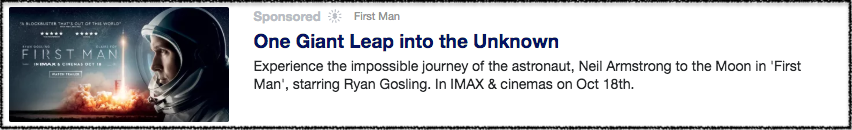}
\caption{A typical Verizon Media native ad taken from Yahoo homepage stream. The ad consists of a title, an image, a description, and a sponsorship notification.}
\label{fig:native ad}
\end{figure}

VZM native serves a daily average of billions of ad impressions to several hundred millions of users world wide, using a sponsorship transparent native ad inventory of several hundred thousands active ads on average. Native ads resemble the surrounding page items, are considered less intrusive to the users, and provide a better user experience in general (see Figure \ref{fig:native ad}).
The online serving system is comprised of a massive Vespa\footnote{Vespa is VZM's elastic search engine solution.} deployment, augmented by ads, budget and model training pipelines. The Vespa index is updated continually with ad and budget changes, and periodically (e.g., every $15$ minutes) with model updates. The VZM native marketplace serves several ad price-types including CPC (cost-per-click), oCPC (optimizing for conversions), CPM (cost-per-thousand impressions), CPV (cost-per-video-view), and also includes RTB (real-time biding) in its auctions. 

\subsection{The {OFFSET} Event-Prediction Algorithm}\label{sec:offset}
The algorithm driving VZM native models is \offset (One-pass Factorization of Feature Sets): a feature enhanced collaborative-filtering (CF)-based ad click-prediction algorithm \cite{aharon2013off}. Focusing on click events, the \textit{predicted click-through-rate} (pCTR) of a given user $u$ and an ad $a$ according to \offset is given by
\begin{equation}\label{eq: pCTR}
    \mathrm{pCTR}(u,a) = \sigma(s_{u,a})\in [0,1]\ ,
\end{equation}
where $\sigma(x)=\left(1+e^{-x}\right)^{-1}$ is the \textit{Sigmoid function}, and 
\begin{equation}\label{eq: score}
	s_{u,a}=b+\nu_{u}^T\nu_{a}\ ,
\end{equation}
where $\nu_{u},\nu_{a}\in \R^D$ denote the user and ad latent factor vectors (LFV) respectively, and $b\in \R$ denotes the model bias. The product $\nu_{u}^T\nu_{a}$ reflects the tendency score of user $u$ towards ad $a$, where a higher score translates into a higher pCTR. To deal with data sparsity issues, the user and ad vectors $\nu_u$ and $\nu_a$ are constructed from individual vectors associated with the user's and the ad's features, such as the user's age and gender, and the ad's category. For the sake of brevity we omit the details of the construction of $\nu_{u},\nu_{a}$ from these vectors, which comprise the model's parameter set $\Theta$, and refer the reader to \cite{aharon2013off}\cite{aharon2017adaptive}.

The parameters $\Theta$ are trained by minimizing the cumulative LogLoss
\[
\mathcal{L}(p, y) = -(1-y) \ln(1-p) - y \ln(p)\ ,
\]
over the training set $\mathcal{T}$ by aiming to minimize the sum of the regularized losses
\[
\sum_{(u, a, y) \in \mathcal{T}} \mathcal{L}(pCTR(u,a), y) + \frac{\lambda}{2} \|\Theta\|^2\ ,
\]
on $\Theta$ using a variant of the AdaGrad \cite{duchi2011adaptive} algorithm, which observes each training sample only once, where $y \in \{0,1\}$ is the click indicator for the event involving user $u$ and ad $a$, and $\lambda$ is the $L2$ regularization parameter. More details of the \offset algorithm are provided in \cite{aharon2013off}\cite{aharon2017adaptive}.

The \offset algorithm also includes an \textit{adaptive online hyper-parameter tuning} mechanism \cite{aharon2017adaptive}. This mechanism takes advantage of the system parallel architecture and strives to tune \offset hyper-parameters (e.g., step size and AdaGrad parameters) to match the varying marketplace conditions, changed by temporal effects and trends. We note that other components of \offset, such as its weighted multi-value feature \cite{arian2019feature}, and similarity weights used for applying ``soft'' recency and frequency rules\footnote{How frequent and how recent a user may be presented with the same ad or campaign.} \cite{aharon2019soft}, are not presented here for the sake of brevity.

\subsection{Accidental Clicks}
In the online advertising jargon, \textit{accidental clicks} (AC) are events where users, browsing some publisher property (e.g., web site, mobile app, etc.), accidentally click on ads, redirected to the advertisers' web page (or \textit{landing page}), and immediately return (or bounce back) to the publisher property. These usually happen due to users' miss interpretation of ads as non-commercial content, or simply due to unintentional clicks, caused mostly due to poor (or even malicious) page design on mobile devices, where clickable elements are placed near ads. Note that clicks are not labeled as accidental and are identifiable mostly via the time period the user spend outside the publisher properties after she clicked an ad, and is referred to as \textit{dwell-time}. Obviously, the click dwell-time feature is only known in retrospect and is not available during serving time.  

In this work we set a time period threshold $\tau_{ac}$ (e.g, $3$ seconds) and use the dwell-time feature to classify the click events in hindsight as follows. All clicks with dwell-time below $\tau_{ac}$ are labeled as \textit{accidental clicks} (AC), otherwise they are labeled as \textit{intentional click} (IC). For more elaborated classification of clicks, the reader is referred to \cite{tolomei2019you}.

\Comment{
\begin{figure}[!t]
\centering
\includegraphics[width=\myWidth\columnwidth]{accClick10.png}
\caption{Percentage of clicks as a function of dwell-time (in sec) from the logged data }
\label{fig:accClickDist}
\end{figure}
}

\section{Related Work}\label{sec:related work}
There are a few published works describing models driving web scale advertising platforms. In \cite{mcmahan2013ad} lessons learned from experimenting with a \textit{large scale logistic regression} model (LSLR) used for CTR prediction by Google advertising system are reported. 
A model that combines decision trees with logistic regression is used to drive Facebook CTR prediction and is reported on in \cite{he2014practical}. The authors conclude that the most important thing for model performance is to have the right features, specifically those capturing historical information about the users or ads.

Recommendation technologies are crucial for CTR prediction, and without them users will find it hard to navigate through the Internet and get what they like. In particular, \textit{collaborative filtering} (CF) in general and specifically \textit{matrix factorization} (MF) based approaches are leading recommendation technologies, where entities are represented by latent vectors and learned by users' feedback (such as ratings, clicks and purchases) \cite{koren2009matrix}. MF-CF based models are used successfully for many recommendation tasks such as movie recommendation \cite{bell2007lessons}, music recommendation \cite{aizenberg2012build}, ad matching \cite{aharon2013off}, and much more. 

VZM has also shared its native ad click prediction algorithm with the community where an earlier version of \textsc{Offset} was presented in \cite{aharon2013off}. A mature version of \textsc{Offset} was presented in \cite{aharon2017adaptive}, where the focus was on the adaptive online hyper-parameter tuning approach, taking advantage of its parallel system architecture. We note that \offset may be seen as a special case of a \textit{factorization machine} \cite{rendle2010factorization}. \offset is also related to \textit{field aware factorization machines} (FFM) \cite{juan2017field}\cite{juan2016field}, and to \textit{field-weighted factorization machines} (FwFM) \cite{pan2018field}. However, it is fundamentally different from those. This is since \offset distinguishes between ad features and user features and allows for triplet-wise dependencies. This comes in contrast to FFM and FwFM, which allow only for pairwise dependencies and treat ad and user features uniformly. Works describing various aspects of VZM native system were also published in recent years. The way frequency capping rules that control how frequent ads are presented to users is presented in \cite{aharon2019soft}. A framework for ``social'' graphs generation and their usage for feature enrichment is described in \cite{arian2019feature}. Ad closes mitigation by fundamental alteration of the GSP auction is presented in \cite{silberstein2020ad}. Finally, \cite{aharon2019carousel} describes how Carousel ads rendering optimization is conducted in VZM native.    

The most relevant work to ours is \cite{tolomei2019you}, which is the only previous work we are aware of, that considers dwell-time modeling, and AC filtering for improving model accuracy. In this work the authors also identified the need to handle ACs for improved click prediction and fair billing of advertisers. In particular, they fit a mixture of three distributions (accidental clicks, short clicks, and long clicks) to dwell-time data logged from Yahoo Gemini native (now VZM native). Using this presentation they set dwell-time thresholds for several traffic segments (see also \cite{yi2014beyond} for another analysis of dwell-time done for Yahoo homepage). Then, they showed that a model trained without ACs (i.e.,  only intentional clicks) provides more accurate CTR predictions than a model trained with all clicks. In-spite of the encouraging offline results, the concept was not incorporated into the VZM native system since it demonstrated under-prediction (reducing the number of auction wins) in online experiments. Moreover, additional efforts (not reported to the community), where ACs were removed from the training and segmented AC biases were added to the predictions prior to serving, also performed poorly, probably due to biases' inaccuracies.

We follow \cite{tolomei2019you} and propose another approach to incorporate the ACs treatment in a way that does not harm the model training on one hand, but also reflects the true CTR of the system on the other, by applying an auxiliary AC prediction model, and using its labels for training the main model. This follows a line of work of \cite{steck2013evaluation}\cite{pmlr-v97-wang19n}\cite{yuan2019improving}, where unobserved data is given labels, potentially by an imputation model, in order to help the main predictive model overcome the inherent selection bias in the available training data.

\section{Click Dwell-Time Analysis}\label{sec: Click Dwell-Time Analysis}

\begin{figure}[!t]
\centering
\includegraphics[width=\columnwidth]{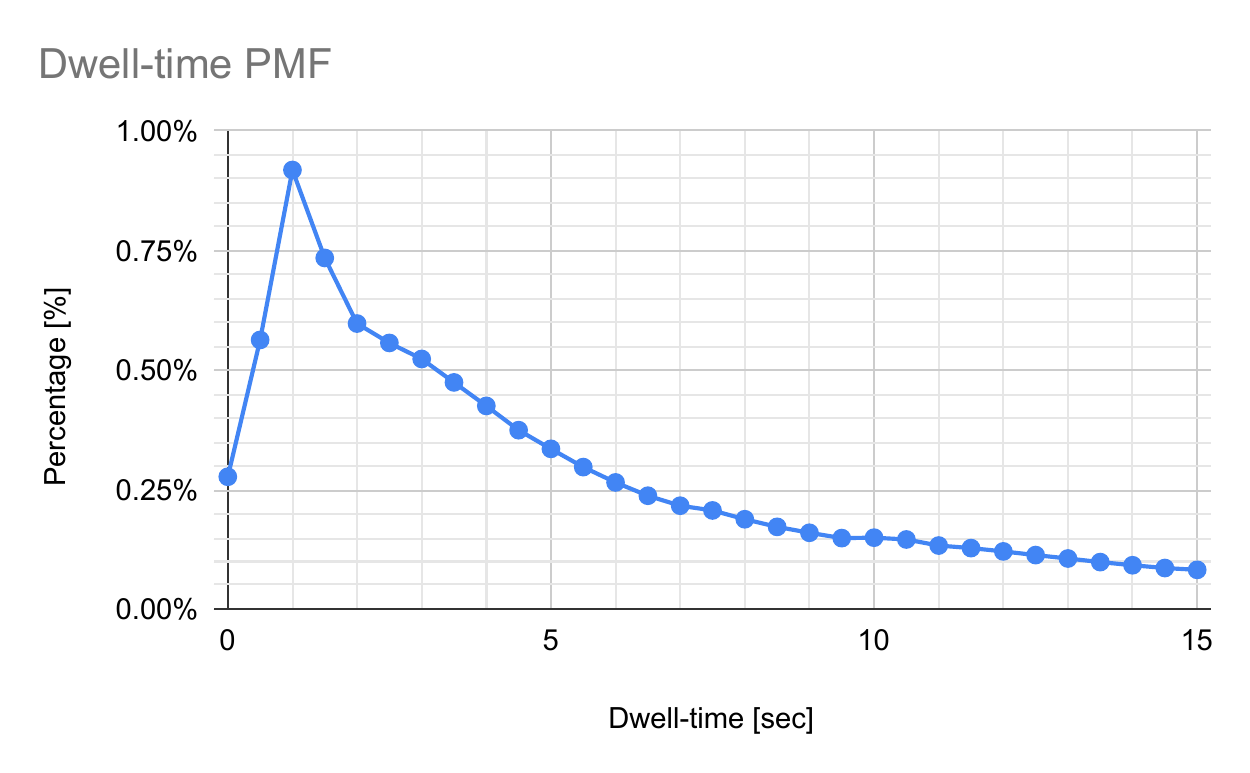}
\caption{Dwell-time probability mass function (PMF).}
\label{fig:DT PDF}
\end{figure}
The following analysis is done for providing insights on ACs and their dwell-time (DT) behaviour, and to support various design decisions regarding the auxiliary AC prediction model (see Section \ref{sec: AC model}).
We have collected click DT data for a period of one week earlier this year. Due to instrumentation issues, DT is available only for $13\%$ of all clicks and partially includes only smartphones, and tablets clicks.

Starting with a global overview, the DT \textit{probability mass function} (PMF) is plotted in Figure \ref{fig:DT PDF}, where all clicks without DT are assumed to have infinite DT (holds for all figures in this section). Setting the AC threshold to $3$ seconds, results in declaring $4.17\%$ of all clicks as accidental.

\begin{figure}[!t]
\centering
\includegraphics[width=\columnwidth]{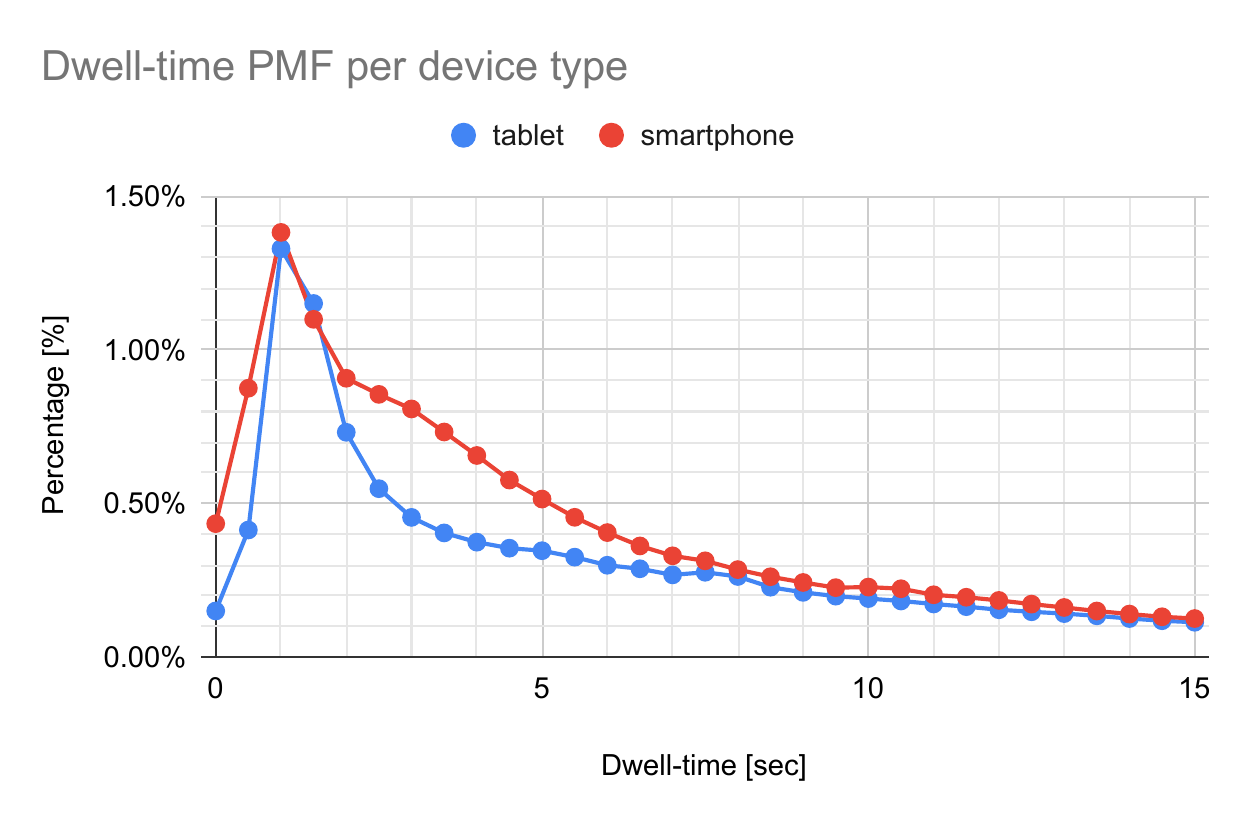}
\caption{Dwell-time PMF per device type.}
\label{fig:DT device PDF}
\end{figure}

In Figure \ref{fig:DT device PDF} the DT PMFs of smartphones and tablets are plotted. Examining the figure it is evident that the statistics are quite different between the device types, where the tablet PMF has a ``havier'' tail. Indeed, features which includes technical details such as device type are useful in AC prediction (see Section \ref{sec: AC model}). Also, setting the $3$ seconds AC threshold we get that $4.77\%$ of clicks are accidental for tablets, while $6.37\%$ of all clicks are accidental on smartphones.

\begin{figure}[!t]
\centering
\includegraphics[width=\columnwidth]{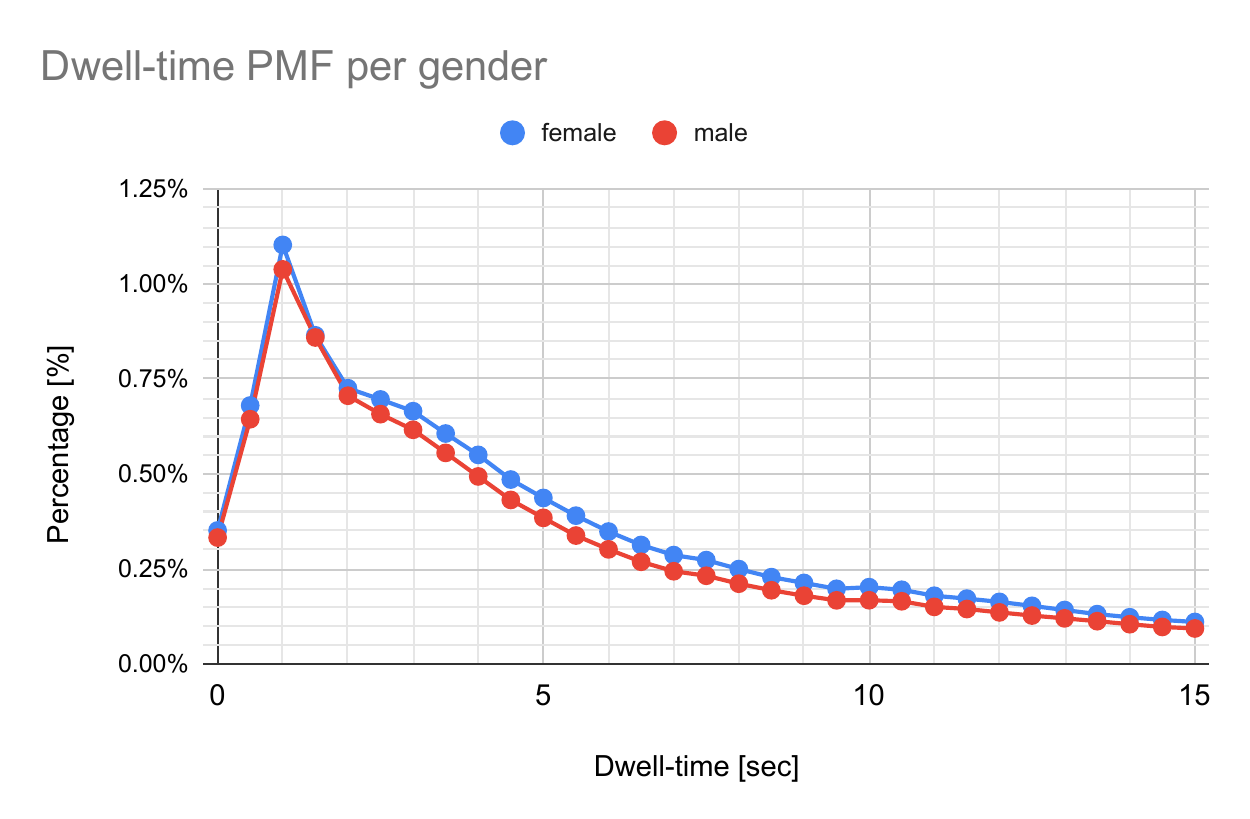}
\caption{Dwell-time PMF per gender.}
\label{fig:DT gender PDF}
\end{figure}

The DT PMF per gender is plotted in Figure \ref{fig:DT gender PDF}. Examining the figure it is observed that female and male statistics are quite similar as expected, demonstrating that demographic features are less useful in AC prediction. Setting the $3$ second threshold we get $5.09\%$ and $4.85\%$ of AC clicks for female and male, respectively.

\begin{figure}[!t]
\centering
\includegraphics[width=\columnwidth]{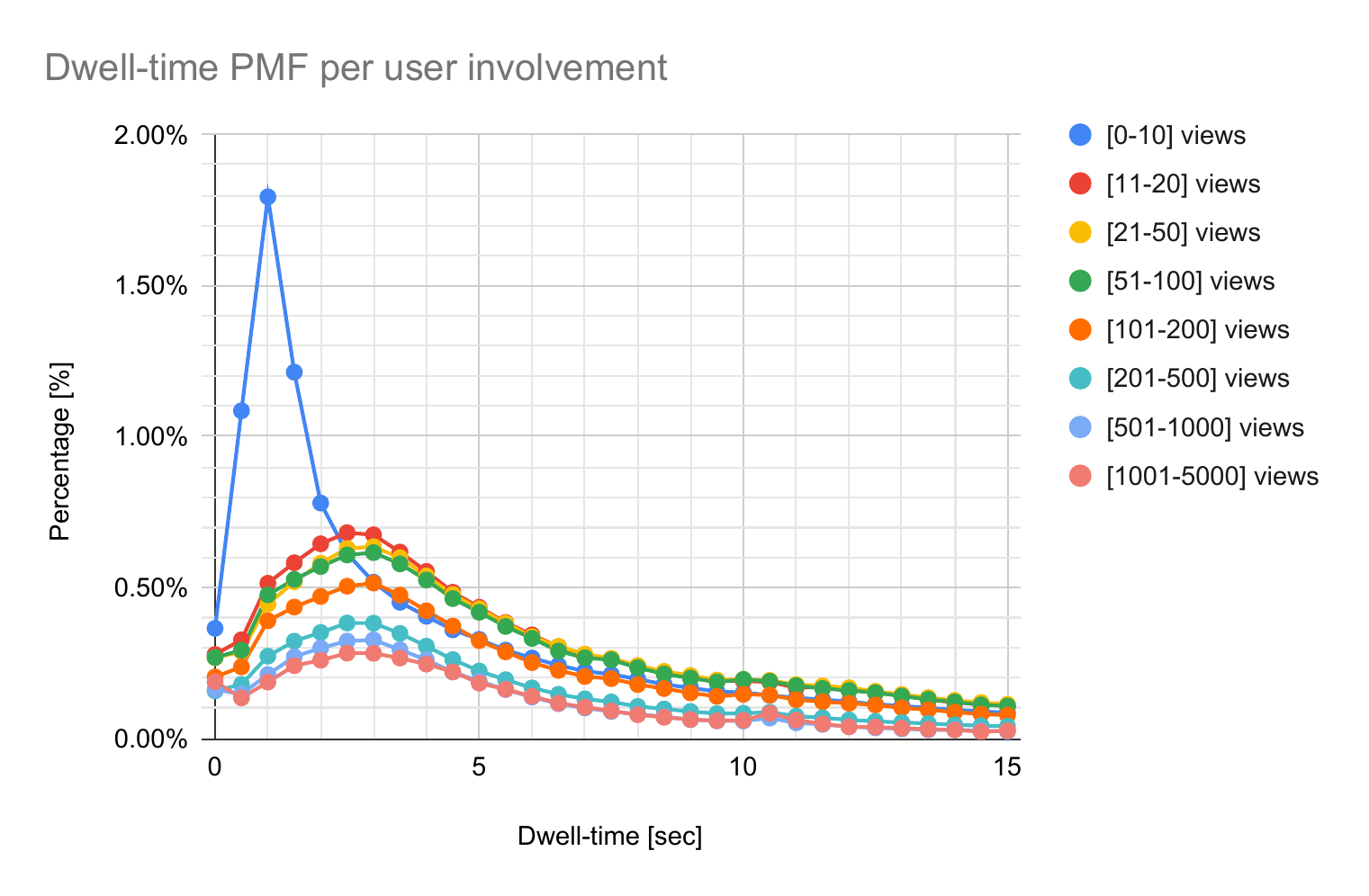}
\caption{Dwell-time PMF per user involvement (number of ads seen during the past month).}
\label{fig:DT ui PDF}
\end{figure}

\begin{small}
\begin{table*}
\centering
\begin{tabular}{c|c|c|c|c|c|c|c|c}
\hline
Ad view bins & [0-10] & [11-20] & [21-50] & [51-100] & [101-200] & [201-500] & [501-1000] & [1001-5000]\\
\hline
AC percentage & $6.37\%$ & $3.70\%$ & $3.37\%$ & $3.36\%$ & $2.76\%$ & $2.05\%$ & $1.74\%$ & $1.58\%$\\
\hline
\end{tabular}
\caption{Accidental click percentage per user involvement (ad views during the past month) bin for $3$ seconds threshold.} \label{tab:AC percentage ui bin}
\end{table*}
\end{small}

User involvement (UI) is a simple yet highly informative user feature, which is defined as the binned number of native ads (or native impressions) seen by the user during a predefined time period (e.g., the past month). In Figure \ref{fig:DT ui PDF}, the DT PMF of several UI bins are plotted. It is observed that the PMF is getting flatter for more involved users. Moreover, the new or occasional visitors of VZM properties (those with fewer views), which are less familiar with its sites, tend more to accidentally click ads. In particular, setting the $3$ seconds AC threshold leads to AC percentage that are presented in Table \ref{tab:AC percentage ui bin}.

\begin{figure}[!t]
\centering
\includegraphics[width=\columnwidth]{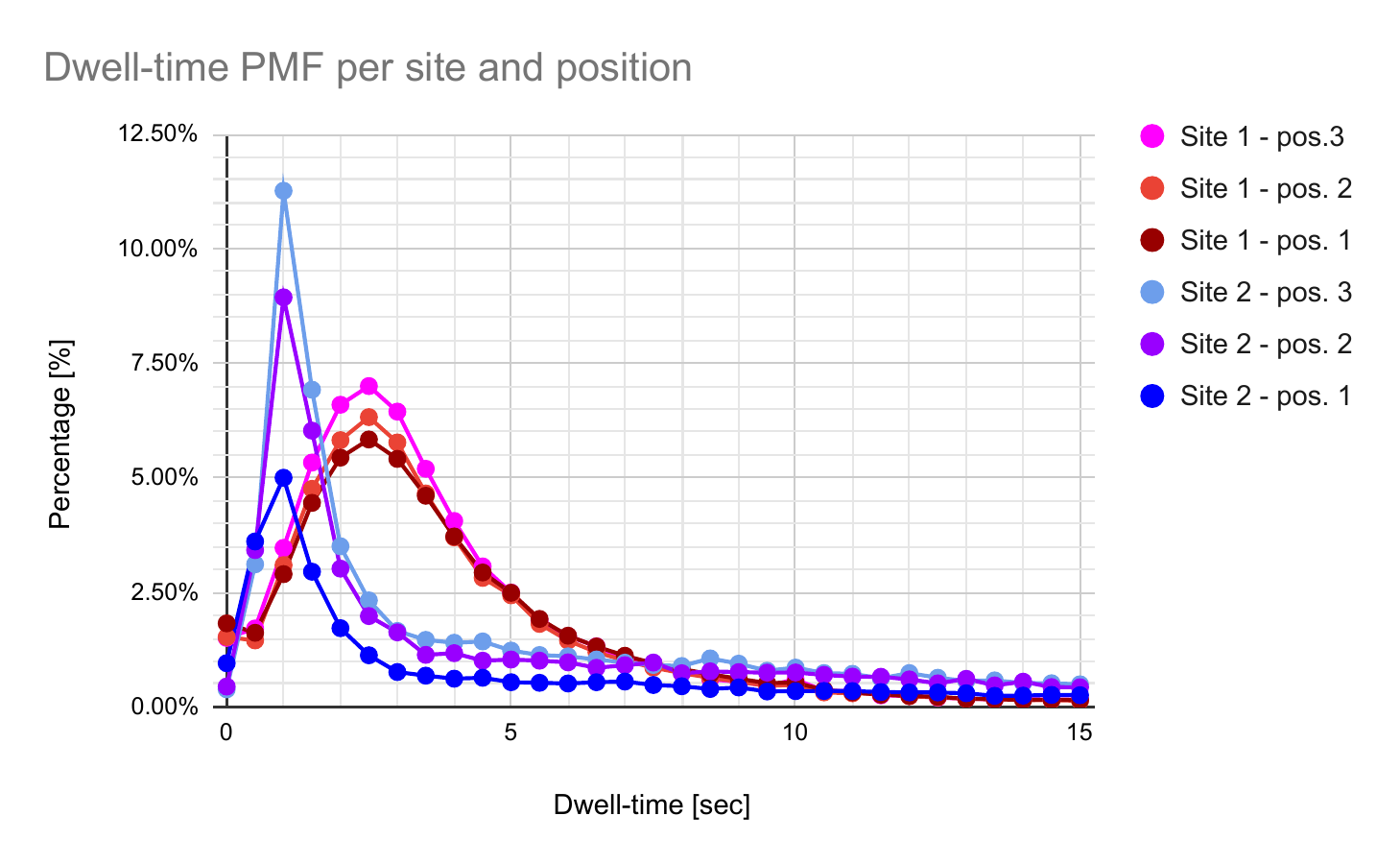}
\caption{Dwell-time PMF per site and positions within the site.}
\label{fig:DT site PDF}
\end{figure}

Many sites include multiple positions for displaying ads\footnote{E.g., a news feed site such as Yahoo homepage may display native ads on the $2$nd and $22$nd slots.} We have chosen two sites and plotted the DT PMF for their first $3$ positions in Figure \ref{fig:DT site PDF}. It is observed that while positions of the same site have similar DT PMF (at least in term of maximal DT probability), the PMF curves rise with position number. This is easily explained by the fact that while same site positions AC probability (or rate) is similar, the top positions that appear near the site header are being seen more and have more clicks in general. 
\begin{figure}[!t]
\centering
\includegraphics[width=\columnwidth]{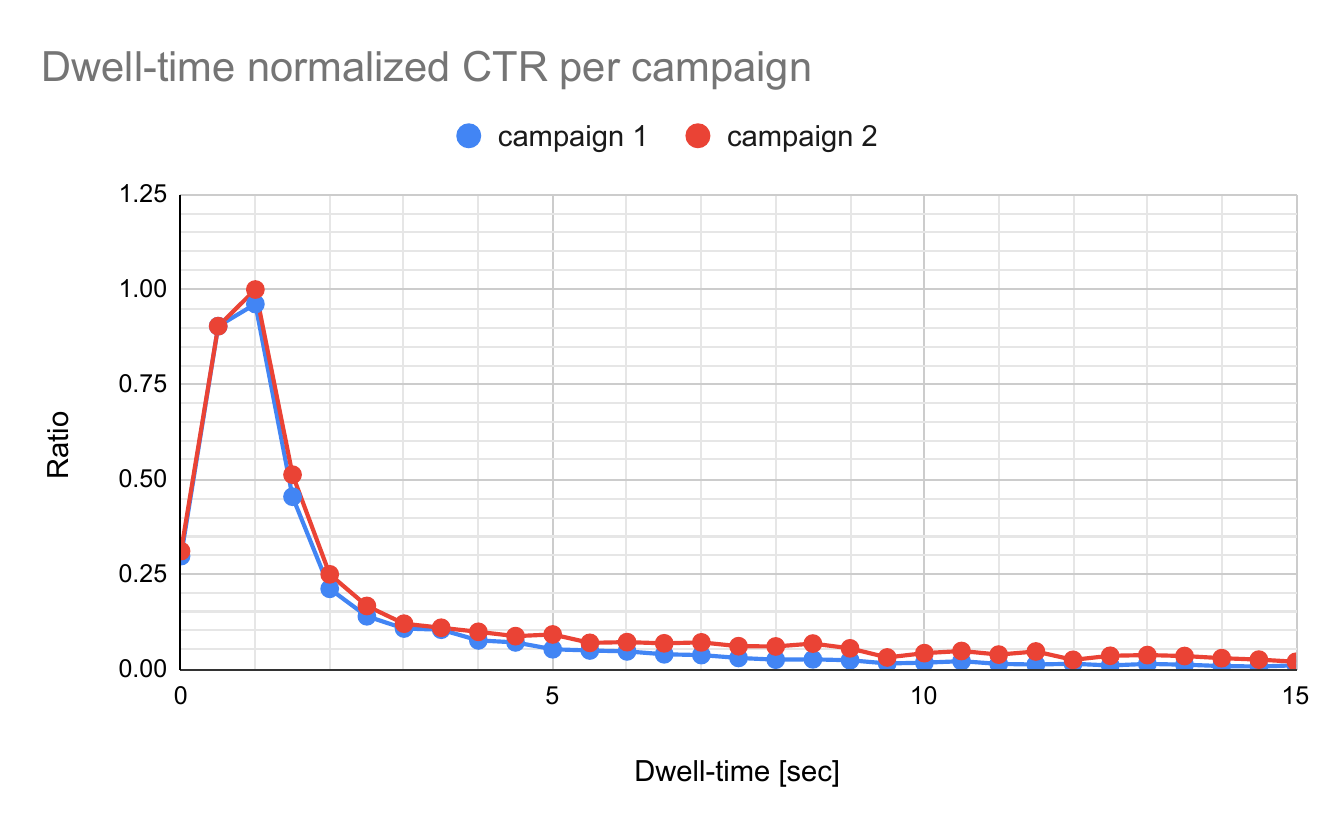}
\caption{Dwell-time normalized CTR for two campaigns measured in a certain site and position.}
\label{fig:DT campaign CTR}
\end{figure}

To complete this section we present evidence that the actual ad or campaign has little effect on the AC rate. To do so we have focused on two active campaigns at a certain site and position. Then, we divided the number of clicks measure for each DT bin by the total number of clicks for each campaign. Due to commercial confidentiality we cannot report CTR values and therefore we normalized all values by the maximal value, and presented the normalized DT CTRs in Figure~\ref{fig:DT campaign CTR}. It is observed that the two curves are almost identical for lower DT values, while their tails are relatively different. In particular, the average normalized CTR of DT values lower than the AC threshold of $3$ seconds, measured for campaign 2, is higher only in $6.1\%$ than that of campaign 1. However, the total CTR of  campaign 2 is higher in $45.5\%$ than that of campaign 1. It is concluded that while the two campaigns may have quite different CTRs they still have similar AC rates. Accordingly, no ad features are used by the auxiliary AC prediction model (see Section \ref{sec: AC model}).

\section{Problem Definition}\label{sec:problemDef}

The problem at hand can be defined as follows.\\

{\bf Increase the system revenue by using the dwell-time feature, available for portion of the training click events.}

\paragraph{Discussion} The essential task of a click prediction model is to learn the user-ad affinities. For this purpose, an AC is harmful. Since the AC would have occurred regardless of which ad was presented, it is intuitively correct to ignore such clicks during training. However, this would cause the model to under-predict, which leads to several unwanted results:
\begin{enumerate}
    \item There are multiple rankers working in parallel to the native ranker\footnote{Such as an e-commerce ranker which suggest specific products  from large catalogs.}, competing over each incoming impression. If one of them is not calibrated, then their predictions aren't comparable, and picking the one with the highest expected revenue may no longer provide the most revenue.
    \item There are sites that apply a floor price, i.e., they ask that the native ranker not return ads unless their expected revenue is above some threshold. An under-predicting model would serve less than it should have on such sites.
    \item There are non-exclusive sites, which decide whether to show the native ad, or choose a different demand source based on expected revenue. Again, an under-predicting model would serve less than it should have on such sites.
\end{enumerate}
Beyond all these reasons for why an under-predicting model would lead to bad outcomes, there is also the fact that we cannot ignore ACs when trying to maximize revenue. For example, assume that a certain site provides only ACs, then, for this site all ads should get the same pCTR and the best ad to present is the one with the highest bid. In general, the model should provide the pCTR of each ad which is a sum of the likelihood of an AC and the likelihood of an intentional one. Filtering ACs will yield a model that ignores the AC component of the pCTR. However, training on all clicks misleads the model about the IC component.

To conclude, it seems that the problem at hand is finding a way to practically remove the ACs from the training set, while not adding a negative bias to the model predictions. Therefore, it is expected that any good modeling approach will provide accurate predictions that sums up closely to the actual number of clicks (both intentional and accidental).

\section{Our Approach}
\label{sec:our approach}
In this Section we describe our unbiased filtering approach for handling ACs.

\begin{figure}[!t]
\centering
\includegraphics[width=\columnwidth]{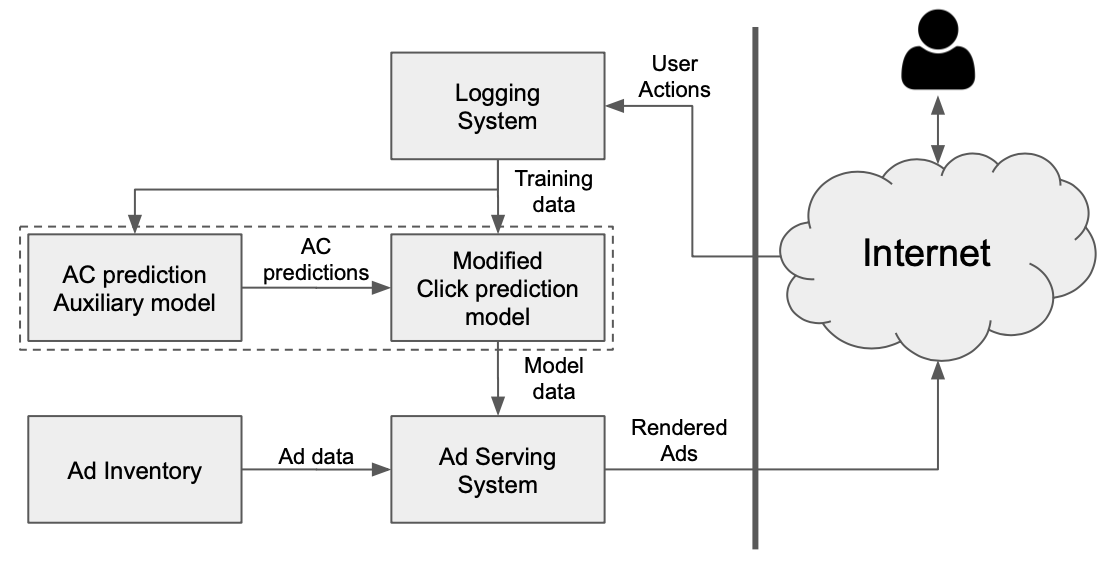}
\caption{High-level system block diagram.}
\label{fig:blobkDiagram}
\end{figure}

\subsection{Overview}
The unbiased filtering approach (see Figure \ref{fig:blobkDiagram} for a high level system block diagram) includes two main components, a modified click prediction model and an auxiliary AC prediction model. The AC model predictions are used as non-binary labels for the modified click model, which uses a binary cross-entropy loss function to optimize its parameters. In the next sub-sections we provide a detailed description of the two models and their interaction, elaborate on their training processes, provide some intuition, and finally discuss some delicate implementation matters.

\subsection{Predicting Accidental Clicks}
\label{sec: AC model}
We use the \offset algorithm (see Section \ref{sec:offset}) to train an auxiliary model to predict the AC probability for every skip event (i.e., impression without click event). We set a threshold and declare all clicks with dwell-time below this threshold as AC, which are used as positive events, while skips are used as negative events. To preserve system resources we train a ``thin'' model with short LFV lengths and a limited number of features. After a feature selection process (partially presented in Section \ref{sec: Offline Evaluation: Auxiliary AC Model}), we focused on the following features:
\begin{itemize}
    \item \textit{User involvement} - the number of ads seen by the user during the past month binned into a few categories (e.g., [0-10], [11-20]\footnote{See Table \ref{tab:AC percentage ui bin} for all bins.}, etc.).
    \item \textit{Tech} - a multi-value feature\footnote{See \cite{arian2019feature} for a detailed description of multi-value features in \offset.} (includes several categorical values for each event) that provides technical details such as the user device type, operating system, browser, Internet provider, and carrier. 
    \item \textit{Site and position} - a categorical feature indicating the position within a certain site. For example, homepage-iOS-1, and homepage-iOS-2, may indicate the first and second ad slots on Yahoo's homepage for iOS devices. 
\end{itemize}
We note that although the \offset framework is described in Section \ref{sec:offset} for user/context and ad features, it can be applied to any feature group that is arbitrary divided into two groups where one group is taking the role of ``user'' features while the other is taking the role of the ``ad'' features. Accordingly, since all the above features are user/context features, we divided them into two groups and applied the \offset algorithm. Experimenting with \offset setting revealed that the best performance in term of LogLoss were achieved when the \textit{user involvement} and \textit{Tech} took the role of ``user'' features, while the \textit{site and position} took the role of ``ad'' features.

\subsection{Modified Click Prediction Algorithm} \label{sec:accoffset}
To train the modified click prediction model we use the basic \offset algorithm (see Section \ref{sec:offset}) with a few modifications. (A) only ICs (i.e., clicks with dwell-time above a predefined time period) are defined as positive events, while skips (i.e., impression with no clicks) and ACs (i.e., clicks with dwell-time below a predefined time period) are defined as negative events; (B) we use non-binary labels, where each positive event is labeled with $1$ and each negative event is labeled with its AC prediction generated by the auxiliary AC model (see Section \ref{sec: AC model}); and (C) to handle non-binary labels we use the more general \textit{binary cross-entropy} loss function (see \cite{mannor2005cross})
\[
\mathcal{L}'(p, \ell) = \ell \ln\left(\frac{\ell}{p}\right) + (1 - \ell) \ln\left(\frac{1 - \ell}{1 - p}\right)\ ,
\]
where $\ell$ is the event's label, and $p$ is the model's prediction. In our case, $p$ is the predicted IC probability, and $\ell$ is $1$ for ICs and the auxiliary AC model prediction otherwise. Note that $\mathcal{L}'$ is non-negative, and zero only if the prediction matches the label, meaning that it is a measure for the deviation of the model's prediction from the label. As with \offset, the modified click model parameters are optimized by minimizing the cross-entropy using SGD like algorithm and applying all other complimentary mechanisms included in the \offset algorithm (see Section \ref{sec:offset}).    
\paragraph{Formal description}
A formal description of the unbiased AC filtering \offset algorithm is elaborated in Algorithm 1.
\input{sfcAlgo}

\paragraph{Why does it work?}
First, let's look from the perspective of a concrete event and understand why the labeling strategy we chose works. Recall, that the cross-entropy loss function measures the deviation of the prediction from the label. Consequently, minimizing the loss means that the label is the ideal prediction: the true probability of a click given an impression. The ideal prediction for an IC is $1$, to reflect the fact that we are sure that the user intentionally clicked on that ad. Both ACs and skips can be treated equivalently, since we assume that these clicks reflect no user intent, and could have been skips (and vice versa). We do not know the true probability of a click for these events, but we have the auxiliary AC prediction model to provide a good guess.

Now consider an entire training set, and assume that the training process indeed minimizes the loss. We will show that the chosen labeling strategy leads to a calibrated model. Denote by $X_c = X_{ac} + X_{ic}$ the random indicators for a click, an AC, and an IC, respectively. Denote also by $p_c$ and $p_{ac}$ the random variable for the prediction of the click and the auxiliary AC models, respectively. Both models are training using the cross-entropy loss, which has a key property shared by all \textit{Bregman divergences} \cite{bregman} - at the optimum, the average label and the average prediction are equal \cite[Theorem 1]{bregman_opt_mean}. In practice we never reach the global optimum, but we assume that we are close, and that the AC model is \textit{nearly} calibrated, namely, $\mathbb{E}[p_{ac}] = \mathbb{E}[X_{ac}] + \varepsilon_{ac}$ for some small $\varepsilon_{ac} > 0$. By the same theorem, the main click prediction model is also nearly calibrated with error $\varepsilon_c$. Since we use auxiliary AC model's predictions as labels of the modified click model, by linearity of expectation we have
\begin{multline*}
\mathbb{E}[p_c] = \mathbb{E}[p_{ac} + X_{ic}] + \varepsilon_{c} \\ = \mathbb{E}[X_{ac} + X_{ic}] + \varepsilon_c + \varepsilon_{ac} = \mathbb{E}[X_c] + \varepsilon_c + \varepsilon_{ac}\ .
\end{multline*}
In other words, the modified click model's average pCTR is also nearly calibrated with error $\varepsilon_c + \varepsilon_{ac}$. 

\subsection{Benefits of Our Approach}
Here is a summary of the main benefits of our approach:
\begin{enumerate}
    \item As mentioned earlier and will be demonstrated in Section \ref{sec:eval}, our approach provides more accurate unbiased click predictions than those of the AC agnostic production model. Moreover, the improved accuracy will be shown to yield significant revenue lifts.
    \item We use a ``thin'' auxiliary AC prediction model with reduced size and fewer features to reduce additional system resources such as memory size and processing power. 
    \item Our approach requires no changes to the serving system. Moreover, the serving system is completely agnostic of ACs and no additional resources are required from it to consume the modified click model. Having a solution that does not requires changes to a mature complex system is most favorable.
\end{enumerate}

\subsection{Implementation Finer Points}
\paragraph{Handling intentional clicks by the auxiliary AC model}
It is not immediately clear what to do with ICs in the auxiliary AC model. Moreover, it might be intuitive to think that ICs should be considered as negative events, as they are impressions that did not yield an AC. Let $T$ be our set of events with $T_{ic},T_{ac},T_n$ being the subsets of ICs, ACs and skips, respectfully. Then the sum of labels observed by the modified click model is $|T_c|+\sum_{e\in T_{ac}\bigcup T_n}\ell(e)$ and we aim to have $\sum_{e\in T_{ac}\bigcup T_n}\ell(e)=|T_{ac}|$. Note that this last requirement does not involve the ICs in any way. We want the sum of predictions of the auxiliary AC model for the skips and ACs events, to be the amount of ACs. If the auxiliary AC model is well-calibrated, this would naturally happen if we were to train on the skips as negative events and the ACs as positive events. It is concluded that the ICs should not be part of the training set of the auxiliary AC prediction model.
\paragraph{Combining with down-sampling}
In the regular \offset training, we reduce processing power and delay by down-sampling negative events (skips) at some rate $R$, while keeping all positive events (clicks). This would naturally lead to high model predictions that can be corrected by adding a factor of $-\log R$ to the model bias (see Eq. \eqref{eq: score}). Naturally, we can either have the auxiliary AC model use the same down-sampling rate, or adjust the labels generated by it to compensate for different sampling rates between the AC model and the modified click model. However, a more subtle issue comes up when considering how to sample the ACs in the modified click model. We train on ACs as skips, so we might think to also down-sample them at the same rate. Using similar notation as in the previous paragraph, we mark $T'_{ac},T'_n$ as the down-sampled sets of ACs and skips, and would want to have $\sum_{e\in T'_{ac}\bigcup T'_n}\ell(e)=|T_{ac}|$. However, the training of the auxiliary AC model only allows for $\sum_{e\in T_{ac}\bigcup T'_n}\ell(e)=|T_{ac}|$, as ACs are positive events for it and so wouldn't be downsampled by its training. It is concluded that we should have that $T'_{ac}=T_{ac}$ and actually not down-sample the ACs in the modified click model.

\section{Performance evaluation}
\label{sec:eval}
In this Section we report the offline performance of the auxiliary AC prediction model, and the offline and online performance of the modified click prediction algorithm. For all cases, we describe the setting and baseline, define the performance metrics, and present the results.

Due to commercial confidentiality matters we report performance lifts only. In addition, for similar reasons and also due to privacy issues our data-sets cannot be published and therefore our experiments cannot be reproduced. 

\subsection{Offline Evaluation: Auxiliary AC Model}
\label{sec: Offline Evaluation: Auxiliary AC Model}

\begin{figure}[tb]
\centering
\includegraphics[width=\columnwidth]{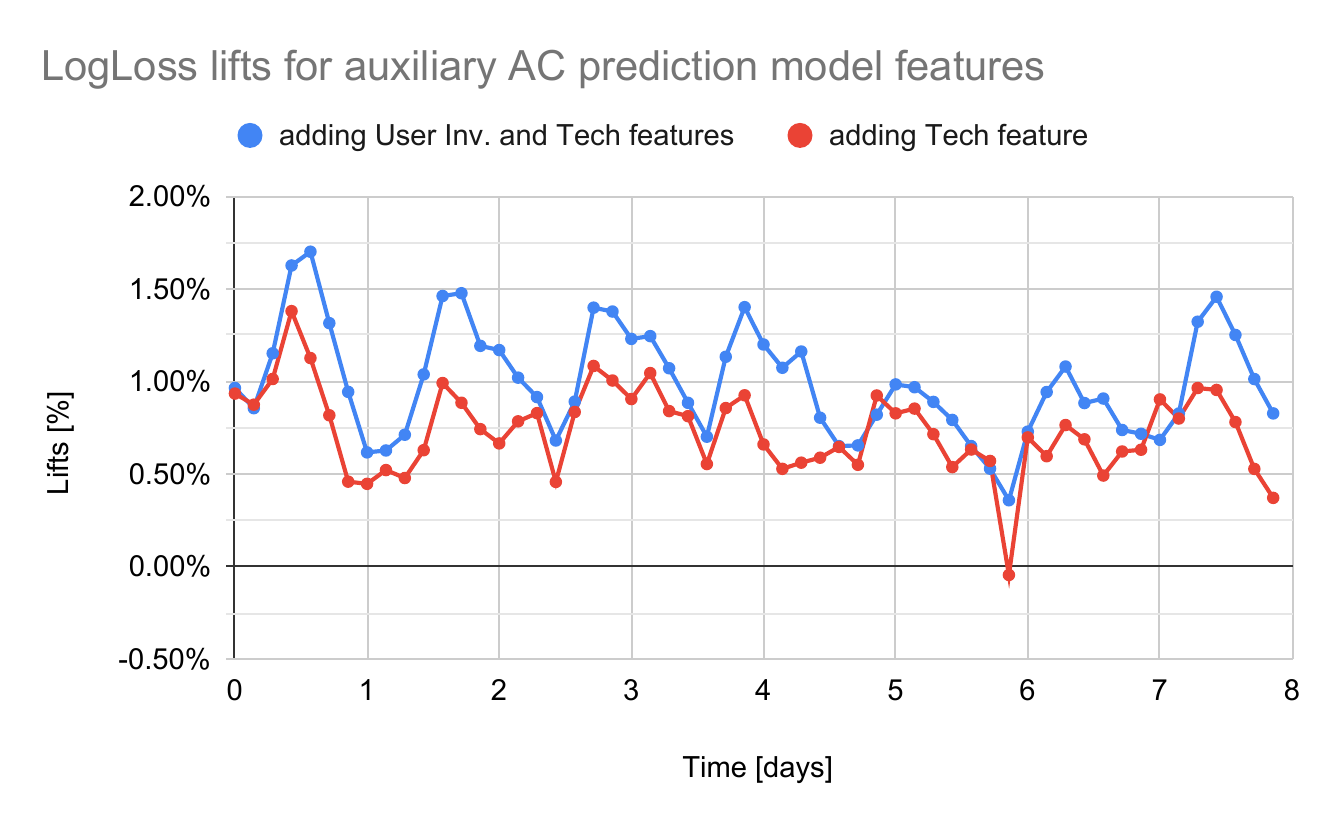}
\caption{LogLoss lifts of the auxiliary AC prediction model, measured for different feature sets for a period of $8$ days  earlier this year.}
\label{fig:offline Ac Model lifts}
\end{figure}

\paragraph{Settings} To evaluate our auxiliary AC prediction model and the individual contribution of each feature, we set the AC threshold to $3$ seconds, and trained three models from ``scratch''\footnote{Training from ``scratch'' means initializing model LFVs in a \textit{Lazy} fashion with Normal random variables $\mathcal{N}(0,\sigma),\ \sigma\ll 1$.} using the \offset algorithm ($o=s=4$, see Section \ref{sec:offset})  and all supporting mechanisms: (1) a baseline model with the \textit{Site and position} feature; (2) same as the baseline model but with the \textit{Tech} feature; and (3) same as model $2$ but with the \textit{User involvement} feature. It is noted that the baseline model may be considered as a weighted ``popularity'' model which uses the average AC rate of every position in every site as its AC prediction.

After training for a few days worth of data, we used the LogLoss metric (defined in the sequel), to measure offline performance, where each event (skip or AC) is used for training the model before being applied to the performance metrics. Hyper-parameters, such as SGD step size and regularization coefficient, are tuned automatically by the adaptive online tuning mechanism included in \offset (see \cite{aharon2017adaptive}).

\paragraph{Results} The LogLoss lifts\footnote{Since lower-is-better with LogLoss metric, the lift is given by $(1-\mathrm{LogLoss_{model}}/\mathrm{LogLoss_{baseline}})\cdot 100$.} of the auxiliary AC prediction model 2 and model 3 over that of the baseline, measured every few hours, are plotted in Figure \ref{fig:offline Ac Model lifts} for a period of $8$ days earlier this year. In particular, we measure average lifts of $0.70\%$ and $0.96\%$ of model 2 and model 3 over the baseline, respectively. Accordingly, adding the \textit{Tech} feature provide $0.70\%$ over the baseline, and adding the \textit{User involvement} provides additional $0.26\%$ over the baseline.

\subsection{Offline Evaluation: Modified Click Model}
\paragraph{Settings}\label{p:setup}
To evaluate offline performance, we train a modified click prediction model as described in Section \ref{sec:accoffset}, and a regular click prediction model which is unaware of ACs, serving as a baseline. Unlike the``thin'' auxiliary AC prediction model, training full size clicks models from ``scratch'' is time consuming. Therefore, we ``seeded''\footnote{``Seeding'' means that we copied a mature model and continue its training using another algorithm, starting with the next data batch.} both modified and baseline models from the production model (which is also unaware of ACs). We use the LogLoss metric (defined next), to measure offline performance, where each impression is used for training the system before being applied to the performance metrics. As with the auxiliary AC model, hyper-parameters, are tuned automatically by the adaptive online tuning mechanism included in \offset (see \cite{aharon2017adaptive}).

Since the two models have different positive and negative label sets, comparing LogLoss results is tricky. To overcome this issue, we measured and compared LogLoss results on traffic segments that do not log dwell-time, so the label sets on those impressions are identical since no ACs are detected there. In particular, we use Yahoo's homepage traffic, which is one of the largest and most stable desktop sites we serve.

As mentioned earlier, the time period threshold we use to define ACs and ICs is $3$ seconds. Setting the threshold to $3$ seconds results in labeling $4.17\%$ of all clicks as accidental. This threshold value was obtained by a simple grid search and yielded the best offline results which are presented in the sequel.

\begin{figure}[!t]
\centering
\includegraphics[width=\columnwidth]{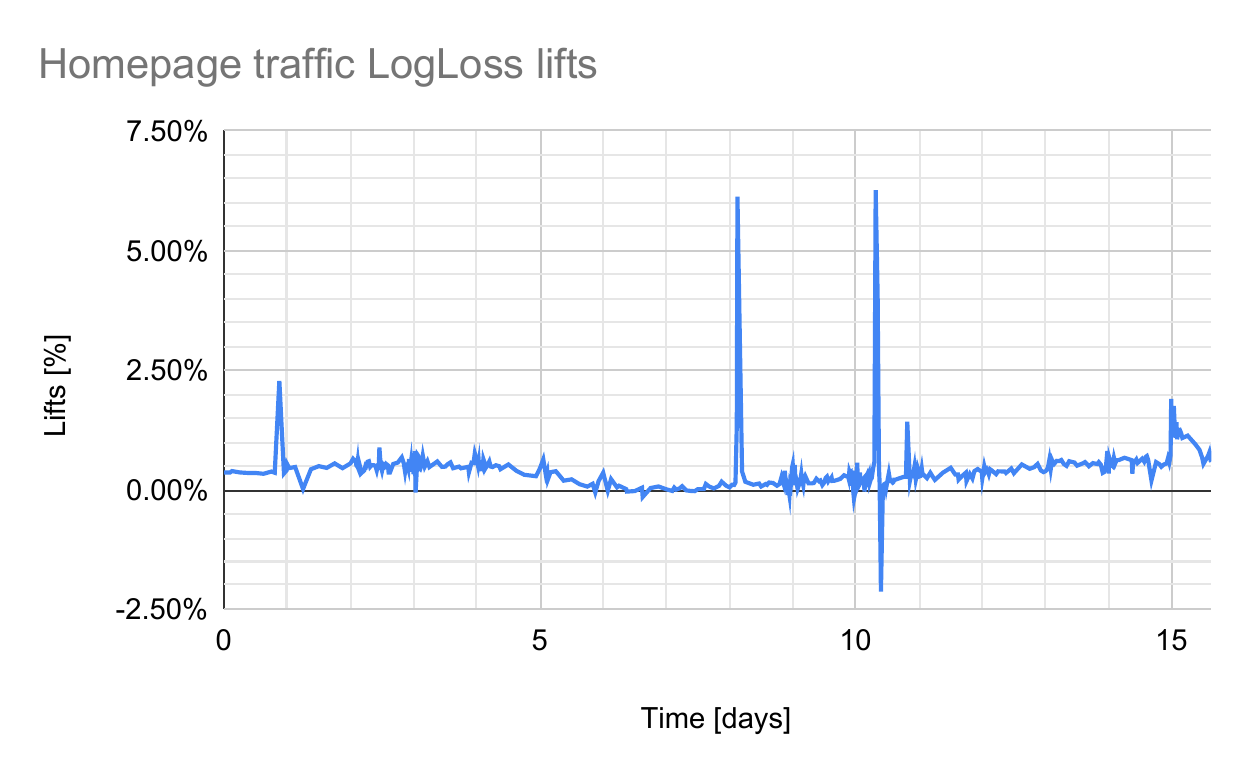}
\caption{LogLoss lifts of the modified model over the production baseline model measured over Yahoo's desktop homepage traffic for a period of $15$ days earlier this year.}
\label{fig:offlineHpstream}
\end{figure}

\paragraph{Evaluation metric}
\begin{description}
\item[LogLoss]
\[
\sum_{(u,a,y)\in \mathcal{T}} \! \! -y \log
pCTR(u,a)-(1-y)\log\left(1-pCTR(u,a)\right)\ ,
\]
where ${\mathcal{T}}$ is a training set and $y \in \{0,1\}$ is the positive event indicator (i.e., click or skip).
\end{description}
We note that the LogLoss metric is used to optimize the regular model's parameters and its regular \offset algorithm hyper-parameters. Moreover, we are using LogLoss despite the fact that we are training the modified click model with cross-entropy loss, since the non-zero skip labels are part of a \textit{training technique}, while for evaluation purposes their true label is still $0$.

\paragraph{Results}
The LogLoss lifts of the modified click model over the baseline model, measured on Yahoo's desktop homepage site traffic every few hours, are plotted for a period of two weeks in Figure \ref{fig:offlineHpstream}. Overall, we measured a hefty average lift of $0.39\%$ in LogLoss. It is noted that such an improvement is not easily achieved for a mature algorithm such as \offset. In addition, since the evaluation is done over billions of impressions, the results are surely statistically significant.

\subsection{Online Evaluation}
\begin{figure}[!t]
\centering
\includegraphics[width=\columnwidth]{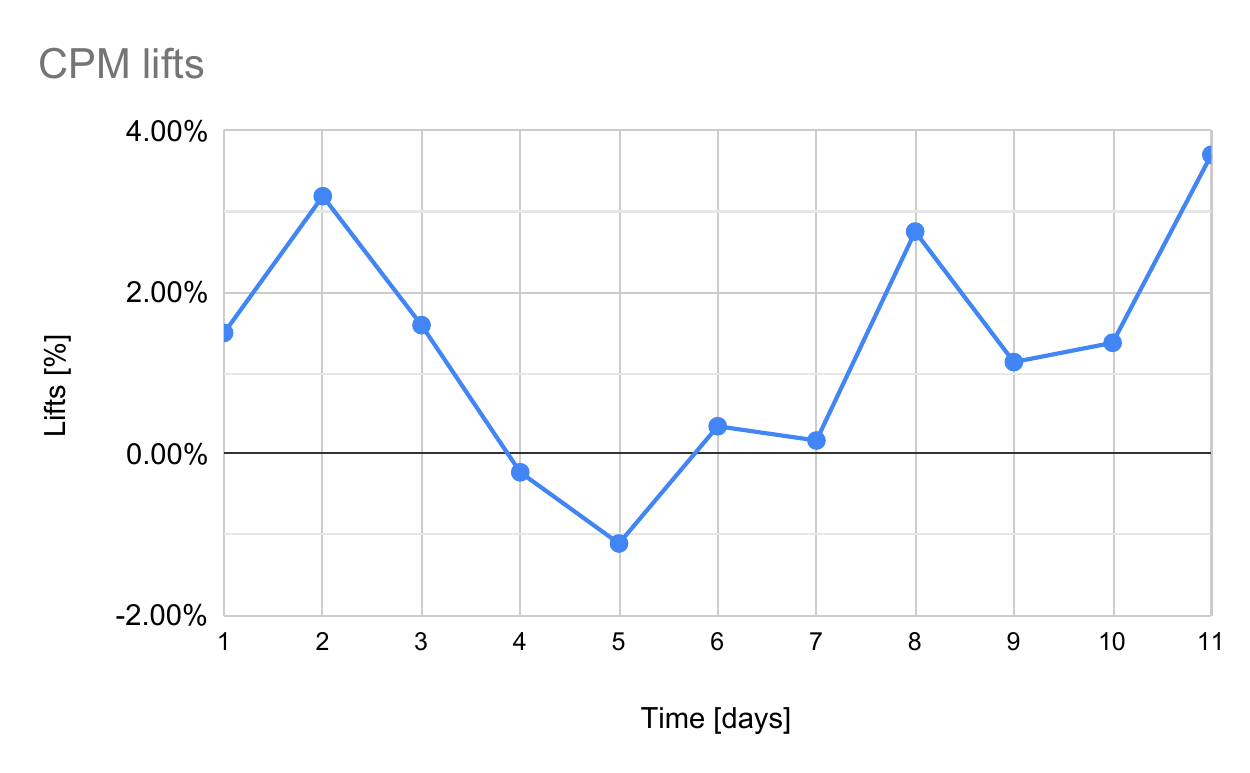}
\caption{Daily CPM lifts of the modified model $1\%$ bucket over the baseline bucket for a period of 11 days earlier this year.}
\label{fig:onlineLift}
\end{figure}

\paragraph{Settings}
To evaluate the online performance of the unbiased AC filtering approach, we launched two online buckets serving $1\%$ of VZM native traffic each (one bucket was serving using the modified click model, while the other was using the baseline production-like model), and measured the revenue lifts in terms of \textit{average cost per thousand impressions} (CPM)\footnote{Since higher-is-better with CPM metric, the lift is given by $(\mathrm{CPM_{AC}}/\mathrm{CPM_{baseline}}-1)\cdot 100$.}. It is noted that since bucket sizes are practically equal, CPM lift is actually revenue lift. 
\paragraph{Results}
The daily CPM lifts of the modified click model bucket when compared to those of the production-like bucket are plotted in Figure \ref{fig:onlineLift} for a period of $11$  days earlier this year. The figure reveals a significant average CPM lift of $1.18\%$. It is noted that such a CPM (or revenue) lift is translated to many millions of USDs in yearly revenue once the solution is deployed to all traffic. As with the offline results, the results here are also statistically significant since evaluation is done over many millions of impressions.

\section{Conclusions and Future Work} \label{sec:conclusions}
In this work we presented a novel approach for handling accidental clicks (AC) using a modified \offset algorithm. The modified algorithm spreads the positive weights of the AC over the negative events using an auxiliary AC prediction model, and uses a binary cross-entropy loss function to train its model parameters. This unbiased AC filtering approach enables us to effectively remove the harmful ACs, that provide little user-ad affiliation information, from the model training process while not hurting the model prediction accuracy. 
After demonstrating promising offline results, the modified model was tested online, serving real VZM traffic and showing a significant $1.18\%$ revenue lifts over a bucket served by an AC agnostic baseline production-like model. It is noted that other solutions were also considered for this problem. The most straightforward solution was to train two models, one for intentional clicks prediction and the other for AC prediction, and let the serving system calculate the combined score of the two models during the auction. This is a valid solution which is expected to demonstrate similar results as the proposed solution. However, it would reduce the query per seconds (QPS) performance of the serving system dramatically while the proposed solution does not require any additional resources from the serving system.   

Future work may include improving the auxiliary AC prediction model accuracy by allocating additional AC relevant features. Another promising direction is to follow \cite{tolomei2019you}, and use segmented thresholds instead of the global $3$ seconds threshold currently used to label ACs.

\section*{Acknowledgement} We would like to thank Natalia Silberstein and Rotem Stram for reviewing the manuscript and providing useful comments.

\bibliographystyle{plain}

\balance
\end{document}

%% file: sfcAlgo.tex
\begin{algorithm}[!htb]
\caption{Unbiased accidental clicks filtering \offset algorithm} \label{algo:frequency modeling}
\textbf{Input:}\\
$\Theta_{c}$ - modified click prediction model\\
$\Theta_{ac}$ - auxiliary AC prediction model\\
$\mathcal{T}$ - training period data\\
$\tau_t$ - AC dwell-time threshold\\
\textbf{Output (updated after each training period):}\\
Updated models $\Theta_{c}$, $\Theta_{ac}$\\
\begin{algorithmic}[1]
\STATE \textbf{AC model $\Theta_{ac}$ update}
\FOR{each event $(u,a,y)\in \mathcal{T}$}
\IF{skip event $y=0$}
    \STATE set label $\ell\coloneqq 0$
    \STATE use \offset to update AC model $\Theta_{ac}$ with $\ell$ 
\ENDIF
\IF{click event $y=1$ and dwell-time $\tau<\tau_t$}
    \STATE set label $\ell\coloneqq 1$
    \STATE use \offset to update AC model $\Theta_{ac}$ with $\ell$ 
\ENDIF
\ENDFOR
\STATE
\STATE \textbf{Modified click model $\Theta_{c}$ update}
\FOR{each event $(u,a,y)\in \mathcal{T}$}
\IF{click event $y=1$ and dwell-time $\tau>\tau_{ac}$}
\STATE set label $\ell\coloneqq 1$
\ELSE
\STATE get AC model $\Theta_{ac}$ prediction $pCTR_{ac}(u)$
\STATE set label $\ell\coloneqq pCTR_{ac}(u)$
\ENDIF
\STATE get modified click model $\Theta_{c}$ prediction $pCTR_{c}(u,a)$
\STATE update click model $\Theta_c$ with label $\ell$ using \offset replacing LogLoss with binary cross-entropy loss
\begin{multline*}
\mathcal{L'}(u,a,\ell) \triangleq \ell\log\left(\frac{\ell}{pCTR_{c}(u,a)}\right)+\\
+(1-\ell) \log \left(\frac{1-\ell}{1-pCTR_{c}(u,a)}\right)+\frac{\lambda}{2}\sum_{\theta\in\Theta_c}\theta^2
\end{multline*}


\ENDFOR
\end{algorithmic}
\end{algorithm} 